\documentclass[runningheads]{llncs}
\usepackage{graphicx}
\usepackage{tikz,pgfplots}
\pgfplotsset{compat=1.18}
\usepackage{amsmath}
\usepackage{amssymb}
\usepackage{booktabs}
\usepackage{xurl}
\usepackage{siunitx}
\usepackage{hyperref}

\makeatletter
\newcommand\notsotiny{\@setfontsize\notsotiny{6}{7}}
\makeatother
\begin{document}
\title{atTRACTive: Semi-automatic white matter tract segmentation using active learning}
\titlerunning{atTRACTive: Semi-automatic tract segmentation using active learning}
%
%
\author{Robin Peretzke\inst{1,2} \and
Klaus Maier-Hein\inst{1,2,3,4,5,6,7} \and
Jonas Bohn\inst{1,4,8} \and
Yannick Kirchhoff\inst{1,6,7} \and 
Saikat Roy\inst{1,7} \and
Sabrina Oberli-Palma\inst{1} \and
Daniela Becker\inst{9,10} \and 
Pavlina Lenga\inst{9} \and
Peter Neher\inst{1,4}
}

%
\authorrunning{R. Peretzke et al.}


%
\institute{Division of Medical Image Computing (MIC), German Cancer Research Center (DKFZ), Heidelberg, Germany 
\and Medical Faculty Heidelberg, Heidelberg University, Heidelberg, Germany 
\and German Cancer Consortium (DKTK), partner site Heidelberg, Germany 
\and National Center for Tumor Diseases (NCT), NCT Heidelberg, a partnership between DKFZ and university medical center Heidelberg 
\and Pattern Analysis and Learning Group, Heidelberg University Hospital, Heidelberg, Germany 
\and HIDSS4Health - Helmholtz Information and Data Science School for Health, Karlsruhe/Heidelberg, Germany \and Faculty of Mathematics and Computer Science, Heidelberg University, Heidelberg, Germany 
\and Faculty of Bioscience, Heidelberg University, Heidelberg, Germany 
\and Department of Neurosurgery, Heidelberg University Hospital, Heidelberg, Germany
\and IU, International University of Applied Sciences, Erfurt, Germany
}

\maketitle              
\begin{abstract}
Accurately identifying white matter tracts in medical images is essential for various applications, including surgery planning and tract-specific analysis. Supervised machine learning models have reached state-of-the-art solving this task automatically. However, these models are primarily trained on healthy subjects and struggle with strong anatomical aberrations, e.g. caused by brain tumors. This limitation makes them unsuitable for tasks such as preoperative planning, wherefore time-consuming and challenging manual delineation of the target tract is typically employed. We propose semi-automatic entropy-based active learning for quick and intuitive segmentation of white matter tracts from whole-brain tractography consisting of millions of streamlines. The method is evaluated on 21 openly available healthy subjects from the Human Connectome Project and an internal dataset of ten neurosurgical cases. With only a few annotations, the proposed approach enables segmenting tracts on tumor cases comparable to healthy subjects (dice~$=0.71$), while the performance of automatic methods, like TractSeg dropped substantially (dice~$=0.34$) in comparison to healthy subjects. The method is implemented as a prototype named atTRACTive in the freely available software MITK Diffusion. Manual experiments on tumor data showed higher efficiency due to lower segmentation times  compared to traditional ROI-based segmentation.

\end{abstract}
\section{Introduction}
Diffusion-weighted MRI enables visualization of brain white matter structures. It can be used to generate tractography data consisting of millions of synthetic fibers or streamlines for a single subject stored in a tractogram that approximate groups of biological axons~\cite{basser2000vivo}. Many applications require streamlines to be segmented into individual tracts corresponding to known anatomy. Tract segmentations are used for a variety of tasks, including surgery planning or tract-specific analysis of psychiatric and neurodegenerative diseases~\cite{berman2009diffusion,mukherjee2006diffusion,mcintosh2008white,wasserthal2020multiparametric}.\\ Automated methods built on supervised machine learning algorithms have attained the current state-of-the-art in segmenting tracts~\cite{berto2021classifyber,wasserthal2018tractseg,zhang2020deep}. Those are trained using various features, either directly from diffusion data in voxel space or from tractography data. Models may output binary masks containing the target white matter tract, or perform a classification on streamline level. However, such algorithms are commonly trained on healthy subjects and have shown issues in processing cases with anatomical abnormalities, e.g. brain tumors~\cite{young2022fibre}. Consequently, they are unsuitable for tasks such as preoperative planning of neurosurgical patients, as they may produce incomplete or false segmentations, which could have harmful consequences during surgery~\cite{yang2021diffusion}. Additionally, supervised techniques are restricted to fixed sets of predetermined tracts and are trained on substantial volumes of hard-to-generate pre-annotated reference data.\\ Manual methods are still frequently used for all cases not yet covered by automatic methods, such as certain populations like children, animal species, new acquisition schemes or special tracts of interests. Experts determine regions of interest (ROI) in areas where a particular tract is supposed to traverse or through which it must not pass, and segmentations can be accomplished either (1) by virtually excluding and maintaining streamlines from tractography according to the defined ROI or (2) by using these regions for tract-specific ROI-based tractography. Both approaches require a similar effort, although the latter is more commonly used. The correct definition of ROIs can be time-consuming and challenging, especially for inexperienced users. Despite these limitations, ROI-based techniques are currently without vivid alternatives for segmenting tracts that automated methods cannot handle.\\ Methods to simplify tract segmentation have been proposed before. Clustering approaches were developed to reduce complexity of large amounts of streamlines in the input data \cite{garyfallidis2012quickbundles,guevara2011robust}. Tractome is a tool that allows interactive segmentation of such clusters by representing them as single streamlines that can interactively be included or excluded from the target tract~\cite{porro2015tractome}. Although the approach has shown promise, it has not yet supplanted conventional ROI-based techniques.\\ We propose a novel semi-automated tract segmentation method for efficient and intuitive identification of arbitrary white matter tracts. The method employs entropy-based active learning of a random forest classifier trained on features of the dissimilarity representation~\cite{olivetti2011supervised}. Active learning has been utilized for several cases in the medical domain, while it has never been applied in the context of tract segmentation~\cite {hao2021transfer,wang2020deep,liu2020semi}. It reduces manual efforts by iteratively identifying the most informative or ambiguous samples, here, streamlines, during classifier training, to be annotated by a human expert. The method is implemented as the tool atTRACTive in MITK Diffusion\footnote{\url{https://github.com/MIC-DKFZ/MITK-Diffusion}}, enabling researchers to quickly and intuitively segment tracts in pathological datasets or other situations not covered by automatic techniques, simply by annotating a few but informative streamlines.

\section{Methods}

\subsection{Binary classification for tract segmentation}
\begin{figure}
\small
    \centering
    \def\svgwidth{0.65\linewidth}
\begingroup%
  \makeatletter%
  \providecommand\color[2][]{%
    \errmessage{(Inkscape) Color is used for the text in Inkscape, but the package 'color.sty' is not loaded}%
    \renewcommand\color[2][]{}%
  }%
  \providecommand\transparent[1]{%
    \errmessage{(Inkscape) Transparency is used (non-zero) for the text in Inkscape, but the package 'transparent.sty' is not loaded}%
    \renewcommand\transparent[1]{}%
  }%
  \providecommand\rotatebox[2]{#2}%
  \newcommand*\fsize{\dimexpr\f@size pt\relax}%
  \newcommand*\lineheight[1]{\fontsize{\fsize}{#1\fsize}\selectfont}%
  \ifx\svgwidth\undefined%
    \setlength{\unitlength}{297.26716665bp}%
    \ifx\svgscale\undefined%
      \relax%
    \else%
      \setlength{\unitlength}{\unitlength * \real{\svgscale}}%
    \fi%
  \else%
    \setlength{\unitlength}{\svgwidth}%
  \fi%
  \global\let\svgwidth\undefined%
  \global\let\svgscale\undefined%
  \makeatother%
  \begin{picture}(1,0.23694456)%
    \lineheight{1}%
    \setlength\tabcolsep{0pt}%
    \put(0,0){\includegraphics[width=\unitlength]{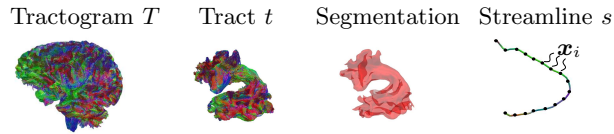}}%
    \put(-0.00029051,0.20266523){\color[rgb]{0,0,0}\makebox(0,0)[lt]{\lineheight{1.25}\smash{\begin{tabular}[t]{l}Tractogram $T$\end{tabular}}}}%
    \put(0.3158997,0.20266523){\color[rgb]{0,0,0}\makebox(0,0)[lt]{\lineheight{1.25}\smash{\begin{tabular}[t]{l}Tract $t$\end{tabular}}}}%
    \put(0.51224914,0.20218232){\color[rgb]{0,0,0}\makebox(0,0)[lt]{\lineheight{1.25}\smash{\begin{tabular}[t]{l}Segmentation\end{tabular}}}}%
    \put(0.78554376,0.20218232){\color[rgb]{0,0,0}\makebox(0,0)[lt]{\lineheight{1.25}\smash{\begin{tabular}[t]{l}Streamline $s$\end{tabular}}}}%
    \put(0.91678639,0.15111817){\color[rgb]{0,0,0}\makebox(0,0)[lt]{\lineheight{1.25}\smash{\begin{tabular}[t]{l}$\boldsymbol x_i$\end{tabular}}}}%
  \end{picture}%
\endgroup%

    \caption{Visualization of a tractogram $T$, a white matter tract $t$ (arcuate fasciculus), its binary segmentation mask and a single streamline $s$ containing  3D points $\boldsymbol{x_i}$.}
    \label{fig:nomenclauture} 
\end{figure}

To create a segmentation of a white matter tract from an individual whole-brain tractogram $T$, streamlines which not belong to this tract must be excluded from the tractography data. This is formulated as a binary classification of a streamline $s\in T$, depending on whether it belongs to the target tract $t$ (see Figure~\ref{fig:nomenclauture} for a brief summary of the nomenclature of this work)
\begin{equation}
\label{eq:1}
    e(s) = \left\{\begin{array}{ll} 1, & s \in t \\
         0, & else\end{array}\right. .
\end{equation}
To perform the classification, supervised models have been trained on various features representing the data. We choose the dissimilarity representation proposed by Olivetti to classify streamlines, which has shown well performance and can be computed quickly for arbitrary data~\cite{berto2021classifyber,olivetti2011supervised}. 
\\
A number of $n$ streamlines, in this case, $n=100$, are used as prototypes forming a reference system of the entire tractogram. A streamline is expressed through a feature vector relative to this reference system. Briefly, a single streamline $s=[\boldsymbol{x_1}, ..., \boldsymbol{x_m}] $, i.e. a polyline containing varying numbers of ordered 3D points $\boldsymbol{x_i}=[x_i, y_i, z_i] \in \mathbb{R}^3, i=1 ... m$, is described by its minimum average direct flip distance $d_{MDF}$ to each prototype~\cite{garyfallidis2015robust}.
The $d_{MDF}$ of a streamline $s_a$ to a prototype $p_a$ is defined as
\begin{equation}
    d_{MDF}(p_a, s_a)=\min((d_{direct}(p_a, s_a), d_{flipped}(p_a,s_a))
\end{equation}
where ${d_{direct}(p_a,s_a)=\frac{1}{m} \sum_{i=1}^m ||\mathbf{x}_i^{p_a} - \mathbf{x}_i^{s_a} ||_2}$ with $m$ being the number of 3D points of the streamlines and ${d_{flip}(p_a,s_a)=\frac{1}{m} \sum_{i=1}^m ||\mathbf{x}_i^{p_a} - \textbf{x}_{m-i+1}^{s_a} ||_2}$.\\ Additionally to $d_{MDF}$, the endpoint distance $d_{END}$ between a streamline and a prototype is calculated, which is equal to $d_{MDF}$, besides, only the start points $x_1$ and endpoints $x_m$ of the streamline and prototype are respected for the calculation~\cite{berto2021classifyber}. Hence, features for a single streamline are represented by a vector twice the number of prototypes. In order to calculate these, all streamlines must have the same number of 3D points and are thus resampled to $m=40$ points.

\subsection{Active learning for tract selection}
\begin{figure}[b!]
    \centering
    \def\svgwidth{\linewidth}
    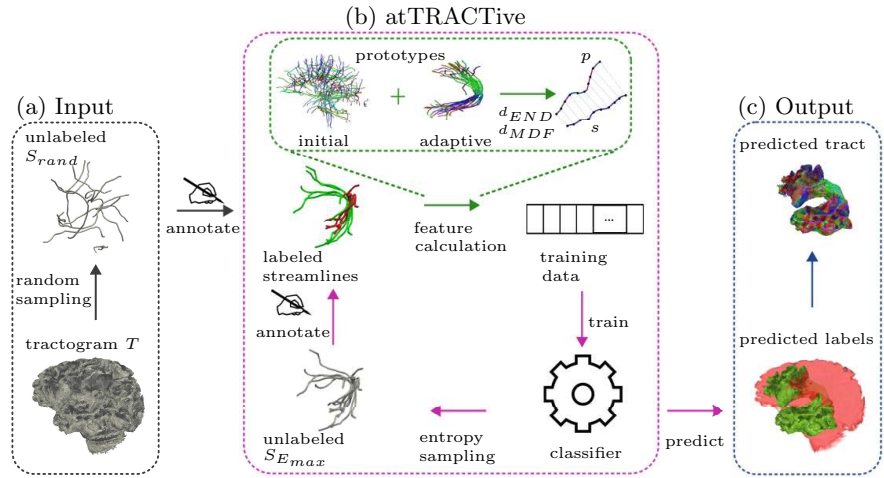
    \caption{Active Learning workflow: Extraction of subset $S_{rand}$ of unlabeled streamlines from the whole-brain tractogram~$T$ for annotation~(a). Initial and adaptive prototypes are used to compute $d_{MDF}$ and $d_{END}$ for each streamline, followed by training the random forest classifier~(b), which predicts on remaining unlabeled streamlines~(c). Entropy reflecting the uncertainty is used to select a new subset $S_{E_{max}}$ to annotate in the subsequent iterations until the human expert is satisfied with the prediction.} 
    \label{fig:workflow} 
\end{figure}
Commonly, for training classifiers, large amounts of annotated and potentially redundant data are used, leading to high annotation efforts and long training times. Active learning reduces both by training machine learning models with only small and iteratively updated labeled subsets of the originally unlabeled data. The proposed workflow is initialized, as shown in Figure~\ref{fig:workflow}~(a), by presenting a randomly selected subset ${S_{rand}=[s_1,..., s_n]}$ of $n=$ streamlines from an individual whole-brain tractogram to an expert for annotation, where ${S_{rand}\subset T}$. Subsequently, the dissimilarity representation is calculated using initially~$100$ prototypes (Figure~\ref{fig:workflow}~(b)), and a classifier is trained, in this case, a random forest. Within completing the training, which takes only a few seconds, the classifier predicts whether the remaining unlabeled streamlines belong to the target tract. Based on the predicted class labels, the target tract is presented (Figure~\ref{fig:workflow}~(c)). Furthermore, the class probabilities $p(s)$ determined by the random forest are used to estimate its uncertainty with respect to each sample by calculating the entropy $E$
\begin{equation}
    \label{eq:2}
    E(s_i) = \sum_{i=1}^n p(s_i) \log p(s_i).
\end{equation}
Next, a subset $S_{E_{max}}$ of streamlines with the highest entropy or uncertainty is selected to be labeled by the expert and is added to the training data (Figure~\ref{fig:workflow}~(c))~\cite{holub2008entropy}. Additionally, these streamlines are utilized as adaptive prototypes until a threshold of $n=100$ adaptive prototype streamlines is reached. Since the model selects ambiguous streamlines in the target tract region, utilizing them as supplementary prototypes improves feature expressiveness in this region of interest.\\ This process is repeated iteratively until the expert accepts the prediction.

\section{Experiments}
\subsection{Data}
The proposed technique was tested on a healthy-subject dataset and on a dataset containing tumor cases. The first comprises 21 subjects of the human connectome project (HCP) that were used for testing the automated methods TractSeg and Classifyber~\cite{wasserthal2018tractseg,berto2021classifyber}. Visual examinations revealed false-negatives in the reference tracts, meaning that some streamlines that belong to the target tract were not included in the reference. These false-negatives did not affect the generation of accurate segmentation masks, since most false-negatives are occupied by true-positive streamlines, but negatively influenced our experiments. To reduce false-negatives, the reference segmentation mask as well as start- and end-region segmentations were used to reassign streamlines from the tractogram using two criteria: Streamlines must be inside the binary reference segmentation~(1) and start and end in the assigned regions~(2). As the initial size of ten million streamlines is computationally challenging and unsuitable for most tools, all tractograms were randomly down-sampled to one million streamlines. We focused on the left optic radiation~(OR), the left cortico-spinal~tract~(CST), and the left arcuate-fasciculus~(AF), representing a variety of established tracts.\\ To test the proposed method on pathological data, we used an in-house dataset containing ten presurgical scans of patients with brain tumors. Tractography was performed using probabilistic streamline tractography in MITK Diffusion. To reduce computational costs, we retained one million streamlines that passed through a manually inserted ROI located in an area traversed by the OR~\cite{tuch2004q}. Subjects have tumor appearance with varying sizes (~($17.87 \pm 12.73 \, \text{cm}^{3}$)) in temporoloccipital, temporal, and occipital regions, that cause deformations around the OR and lead to deviations of the tract from the normative model.

\subsection{Experimental setup}
 To evaluate the proposed method, we conducted two types of experiments. Manual segmentation experiments using an interactive prototype of atTRACTive were initiated on the tumor data (holistic evaluation). Additionally, reproducible simulations on the freely available HCP and the internal tumor dataset were created (algorithmic evaluation). In order to mimic expert annotation during algorithmic evaluation, class labels were assigned to streamlines using previously generated references. The quality of the predictions was measured by calculating the dice score of the binary mask. The code used for these experiments is publicly available\footnote{\url{https://github.com/MIC-DKFZ/atTRACTive_simulations}}.\\ For the algorithmic evaluation, the initial training dataset was created with $20$ randomly selected streamlines from the whole-brain tractogram, which have been shown to be a decent number to start training. Since some tracts contain only a fraction of streamlines from the entire tractogram, it might be unlikely that the training dataset will contain any streamline belonging to the target tract. Therefore, two streamlines of the specific tract were further added to the training dataset, and class weights were used to compensate for the class unbalance. According to Figure~\ref{fig:workflow}, the dissimilarity representation was determined, the random forest classifier was trained and the converged model was used to predict on the unlabeled streamlines and to calculate the entropy. In each iteration, the ten streamlines with the highest entropy are added to the training dataset, which has been determined to be a good trade-off between annotation effort and prediction improvement. The process was terminated after $20$ iterations, increasing the size of the training data from $22$ to $222$ out of one million streamlines.\\ The holistic evaluation was conducted with equal settings, except that the workflow was terminated when the prediction matched the expectation of the expert. To ensure that the initial dataset $S_{rand}$ contained streamlines from the target tract, the expert initiated the active learning workflow by defining a small ROI that included fibers of the tract. $S_{rand}$ was created by randomly sampling only those streamlines that pass through this ROI. To allow comparison between the proposed and traditional ROI-based techniques, the OR of subjects from the tumor dataset were segmented using both approaches by an expert familiar with the respective tool, and the time required was reported to measure efficiency.\\ Note, in all experiments, the classifier is trained from scratch every iteration, prototypes are generated for each subject individually, and the classifier predicts on data from the same subject it is trained with, as it performs subject-individual tract segmentation and is not used as a fully automated method. To ensure a stable active learning setup that generalizes across different datasets, the whole method was developed on the HCP and applied with fixed settings to the tumor data~\cite{luth2023toward}. 
 
\subsection{Results}
\begin{figure}
    \begin{center}
    \resizebox{.65\linewidth}{!}{\input{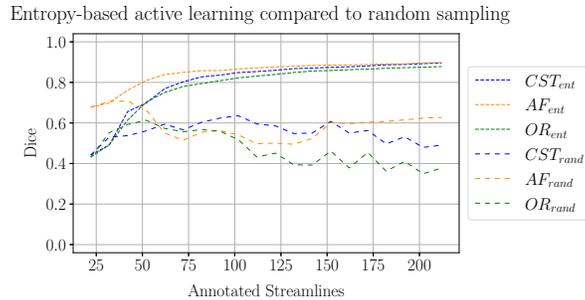}}
    \end{center}
    \caption{Dice score of entropy-based active learning of the HCP dataset of the OR, CST, and AF ($[Tract]_{ent}$) compared to random sampling ($[Tract]_{rand})$}
    \label{fig:quantitative}
\end{figure}

\begin{figure}
\notsotiny
    \centering
    \def\svgwidth{\linewidth}
    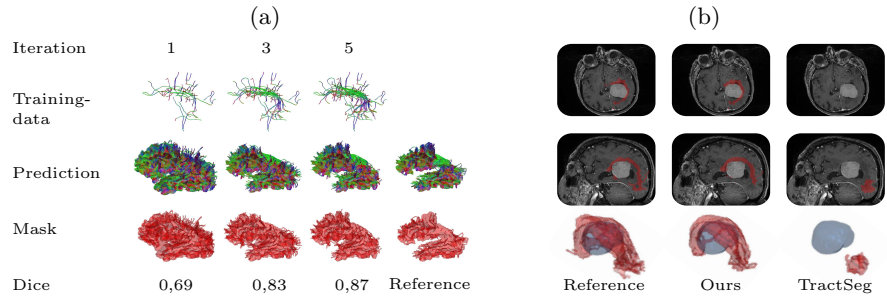
    \caption{Training data, prediction, segmentation mask, and dice-score after first, third, and fifth iteration of active learning with the reference of the AF~(a) and tumor (blue) and OR segmentation (red) with reference, active learning outcome after the third iteration and TractSegs output~(b).} 
    \label{fig:qualitative}
\end{figure}

In Table~\ref{tab:res_HCP}, the dice score of the active learning simulation on the HCP and tumor data after the fifth, tenth, and twentieth iterations are shown and compared with outcomes of Classifyber and TractSeg. Results for the HCP data were already on par with the benchmark of automatic methods between the fifth and tenth iterations. On the tumor data, the performance of the proposed method remains above $0.7$ while the performance of TractSeg drops substantially. Furthermore, Classifyber does not support the OR and is therefore not listed in Table~\ref{tab:res_HCP}.
\begin{table}
\setlength{\tabcolsep}{5pt} 
\renewcommand{\arraystretch}{1} 
    \centering
        \begin{tabular}{l| c c c|c}
         & \multicolumn{3}{c|}{HCP dataset} & Tumor dataset \\
        \hline
        Tract   &   CST &   AF &  OR &  OR\\
        \hline
        Classifyber \cite{berto2021classifyber} & $0.86 \pm 0.01$ & $0.84 \pm 0.03$  & - & -\\
           TractSeg \cite{wasserthal2018tractseg} & $0.86 \pm 0.02$ & $0.86 \pm 0.03$ & $0.83 \pm 0.02$ & $0.34 \pm 0.19$ \\
           Active Learning\textsubscript{it=5} & $0.83 \pm 0.05$ & $0.86 \pm 0.03$ & $0.79 \pm 0.11$ & $0.71 \pm 0.06$ \\
             Active Learning\textsubscript{it=10} & $0.88 \pm 0.03$ & $0.88 \pm 0.02$ & $0.85 \pm 0.05$ & $0.72 \pm 0.05$ \\
             Active Learning\textsubscript{it=20} & $\mathbf{0.90 \pm 0.03}$ & $\mathbf{0.90  \pm 0.02}$ & $\mathbf{0.88 \pm 0.02}$ & $\mathbf{0.73 \pm 0.08}$\\
        \hline
        \end{tabular}
        \vspace{3pt}
    \caption{Dice score of TractSeg and Classifyber on both datasets compared to entropy-based active learning after the fifth, tenth, and twentieth iteration.}
    \label{tab:res_HCP}
\end{table}\\
Figure~\ref{fig:quantitative} depicts the quantitative gain of active learning on the three tracts of the HCP data and compares it to pure random sampling by displaying the dice score depending on annotated streamlines. While active learning leads to an increase in the metric until the predictions at around five to ten iterations show no meaningful improvements, the random selection does not improve overall.
\\
Qualitative results of the algorithmic evaluation of the AF of a randomly chosen subject of the HCP dataset are shown in Figure~\ref{fig:qualitative}~(a). Initially, the randomly sampled streamlines in the training data are distributed throughout the brain, while entropy-based selected streamlines from subsequent iterations cluster around the AF. The prediction improves iteratively, as indicated by a rising dice score. When accessing qualitative results of the pathological dataset visual inspection revealed particularly poorly performance of TractSeg in cases where OR fibers were in close proximity to tumor tissue, leading to fragmented segmentations, while complete segmentations were reached with active learning even for these challenging tracts after a few iterations, as shown in Figure~\ref{fig:qualitative}~(b).\\ The initial manual experiments with atTRACTive were consistent with the simulations. The prediction aligned with the expectations of the expert at around five to seven iterations taking a mean of 4,5 minutes, while it took seven minutes on average to delineate the tract with ROI-based segmentation. During the iterations, streamlines around the target tract were suggested for labeling, and the prediction improved. Visual comparison yielded more false-positive streamlines with the ROI-based approach while atTRACTive created more compact tracts.

\section{Discussion}
Active learning-based white matter tract segmentation enables the identification of arbitrary pathways and can be applied to cases where fully automated methods are unfeasible. In this work, algorithmic evaluation as well as the implementation of the technique into the GUI-based tool atTRACTive
including further holistic manual experiments were conducted.\\ The algorithmic evaluation yielded consistent results from the fifth to the tenth iterations on both the HCP and tumor datasets. As expected, outcomes obtained from the tumor dataset were not quite as good as those of the HCP dataset. This trend is generally observed in clinical datasets, which tend to exhibit lower performance levels compared to high-quality datasets, which could be responsible for the decline in the results. Preliminary manual experiments with atTRACTive indicated active learning to have shorter segmentation times compared to traditional ROI-based techniques. These experiments are in line with the simulations as the generated tracts matched the expectations of the expert after around five to seven iterations, meaning that less than a hundred out of million annotated streamlines are required to train the model. Enhancements to the usability of the prototype are expected to further improve efficiency. A current limitation of \mbox{atTRACTive} is the selection of the initial subset, based on randomly sampling streamlines passing through a manually inserted ROI. This approach does not guarantee that streamlines of the target tract are included in the subset. In that case, the ROI has to be replaced or $S_{rand}$ needs to be regenerated.\\ Future analyses, evaluating the inter- and intra-rater variability compared to other interactive approaches, will be conducted on further tracts. For selected scenarios, the ability of the classifier to generalize by learning from previously annotated subjects will be investigated, which may even allow to train a fully automatic classifier for new tracts once enough data is annotated. To further optimize the method, the feature representation or sampling procedure could be improved. Uncertainty sampling may select redundant streamlines due to similar high entropy values. Instead, annotating samples with high entropy values being highly diverse or correcting false classifications could convey more information.\\ By introducing active learning into tract segmentation, we provide an efficient and intuitive alternative compared to traditional ROI-based approaches. \\\mbox{atTRACTive} has the potential to interactively assist researchers in identifying arbitrary white matter tracts not captured by existing automated approaches.

\bibliographystyle{splncs04}
\bibliography{mybibliography.bib}

\end{document}